\documentclass[twocolumn,showpacs,amsmath,amssymb]{revtex4}

\usepackage{amsmath,amssymb,latexsym,graphics,epsfig}

\begin{document}

\title{Complex motion of precipitation bands}

\author{Istv\'an Lagzi}
\email{lagzi@vuk.chem.elte.hu}
\affiliation{Institute of Chemistry, E\"otv\"os University, 1117 Budapest, P\'azm\'any s\'et\'any 1/a, Hungary,}
\author{P\'eter P\'apai}
\email{papai@general.elte.hu}
\author{Zolt\'an R\'acz}
\email{racz@general.elte.hu}
\affiliation{Institute for Theoretical Physics - HAS, E\"otv\"os University, 1117 Budapest, P\'azm\'any s\'et\'any 1/a, Hungary}

\begin{abstract}
Formation and dynamics of an Al(OH)$_3$ precipitation ring is
studied by diffusing NaOH into a gel containing AlCl$_3$. Limited
feeding of the outer electrolyte (NaOH) is found to yield an
intricate ring-dynamics which involves stopping and reversal of the
direction of motion of the precipitation ring, and evolution into
stationary multi-ring structures. A model of the ring-dynamics is
developed by combining a phase separation scenario for the
precipitation with the redissolution (complex formation) of the
precipitate in the excess of the outer electrolyte.
\end{abstract}
\maketitle

\section{Introduction}
Precipitation patterns formed in the wake of moving reaction fronts
have been studied for a long time \cite{lie}. Recent interest in
these structures stems from possible relevance to engineering
mesoscopic and microscopic patterns
\cite{Giraldo2000,Lebedeva2004a,Lebedeva2004b,Fialkowski2005,Grzybowski2005}.
In contrast to removing material to construct a structure (top-down
processing), controlled precipitation is suggested as means to
generate a prescribed bulk design (bottom-up processing). In order
to understand how to guide and locate precipitation regions, it is
instrumental to investigate cases where, in addition to
precipitation in the reaction zone, redissolution in the wake of the
moving front also takes place
\cite{Zrinyi91,Das89,Das91,Sultan96,Das97,Sultan02,Lagzi03,Volford06,George2005}.
The resulting virtual motion of a precipitation pulse (or more
complicated pattern) is easily visualized and its dynamics can be
studied in detail.

In a usual setup for observing moving precipitation pulses, the
inner electrolyte is distributed homogeneously in a gel column and
the outer electrolyte diffuses into this medium (typical examples of
inner/outer electrolyte pairs are Hg$^{2+}$/I$^-$,
Co$^{2+}$/NH$_4$OH, Cr$^{3+}$/OH$^-$). The precipitate (HgI$_2$,
Co(OH)$_2$, Cr(OH)$_3$) forms in the reaction zone which moves
diffusively along the column. In the wake of this front, the excess
of the outer electrolyte consumes the precipitate and forms a
complex ([HgI$_4]^{2-}$, [Co(NH$_3)_6]^{2+}$, [Cr(OH)$_4]^{-}$). In
cases when the gel, the reagents, and the complex are transparent,
while the precipitate is not, the position of the precipitation
pulse (band) is easily monitored.

Our aim here is to study the dynamics of precipitation pulses in
radially symmetric, two-dimensional setup with limited feeding of
the outer electrolyte. From experimental side, we present a
quantitative description of the radial motion of precipitation
pulses with emphasis on the stopping of the band and on the reversal
of the radial motion with the ensuing multiplication of the
precipitation bands. From theoretical side, we generalize the so
called phase separation theory of Liesegang phenomena \cite{ModelB},
and show that the novel aspects of the experiments (reversal of
front motion and the multiplication of bands) can be described well
but problems arise when such detail as the time-evolution of the
width of the precipitation band is considered.

\section{Experiments}
\label{Exp}
The evolution of the precipitation pulse is observed in
the following reaction scheme \cite{Volford06}:
\begin{eqnarray}
\textrm {Al}^{3+}(\textrm {aq}) + 3 \textrm {OH}^{-}(\textrm {aq}) &\rightarrow& \textrm {Al(OH)}_3(\textrm{s})\nonumber\\
\textrm{Al(OH)}_3(\textrm{s}) + \textrm{OH}^{-}(\textrm{aq})
&\rightarrow& [\textrm{Al(OH)}_4]^{-}(\textrm{aq}).\nonumber
\end{eqnarray}
Agarose (Reanal) was dissolved in distilled water to produce 1\%
solution. It was continuously stirred and heated up to 90 $^o$C. The
clear solution was mixed with the given amount of AlCl$_3$*6H$_2$0
(Reanal), and this solution was poured into a Petri dish to obtain a
uniformly thick gel (3.2 mm). After polymerization, a circular hole
of radius $R$ is cut out at the center of the gel, and the outer
electrolyte of fixed concentration ($a_0=2.50$M) was placed into
this reservoir. No feeding of the  outer electrolyte  (usual
boundary  conditions  in  many experiments  and  simulations) was
allowed. The parameters varied in the experiments were the
concentration of the inner electrolyte ($b_0$) and  the radius  of
the  reservoir ($R$). When changing $R$, the volume $V_0$ of the
outer electrolyte was changed proportionally to $R^2$.

Shortly after the outer electrolyte is placed into the reservoir, a
white  precipitate  can  be observed  at the gel-reservoir
interface. Next, the  thin precipitation  ring  detaches from the
interface (due to redissolution of the precipitate in the wake of
the outward moving reaction front). The precipitation band  is  well
visible because  both  the  aluminum chloride  and its hydroxo
complex are colorless in contrast to the white precipitate,
Al(OH)$_3$.  At early stages, the increase of radius of the pulse
$r_f(t)$ is proportional to the square root of time $r_f(t)-R\sim
\sqrt{t}$, indicating that the front motion is driven by diffusion
of the  invading  electrolyte. Then the  front  motion slows down
(faster than in case of continuous feeding) and,  finally,  the
front  stops (Figure  1a). The total increase in the radius
$r_f(\infty)-R$ depends on the concentrations of the  inner- and
outer   electrolytes, and on radius of the reservoir.  The value of
$r_f(\infty)-R$ can  be easily estimated  from the mass conservation
law.

\begin{figure}[htb]
\includegraphics[width=8.0cm]{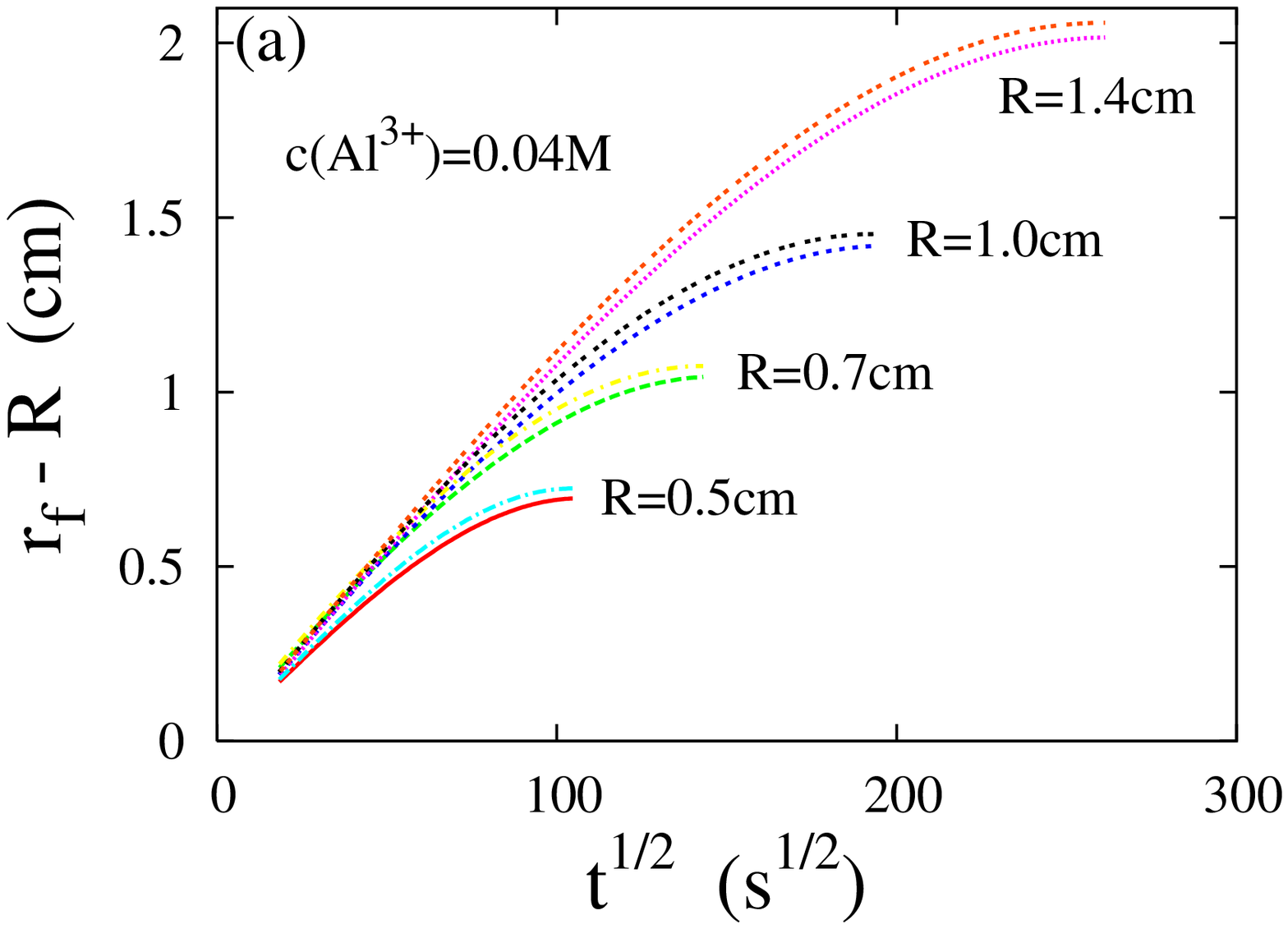}
\includegraphics[width=8.0cm]{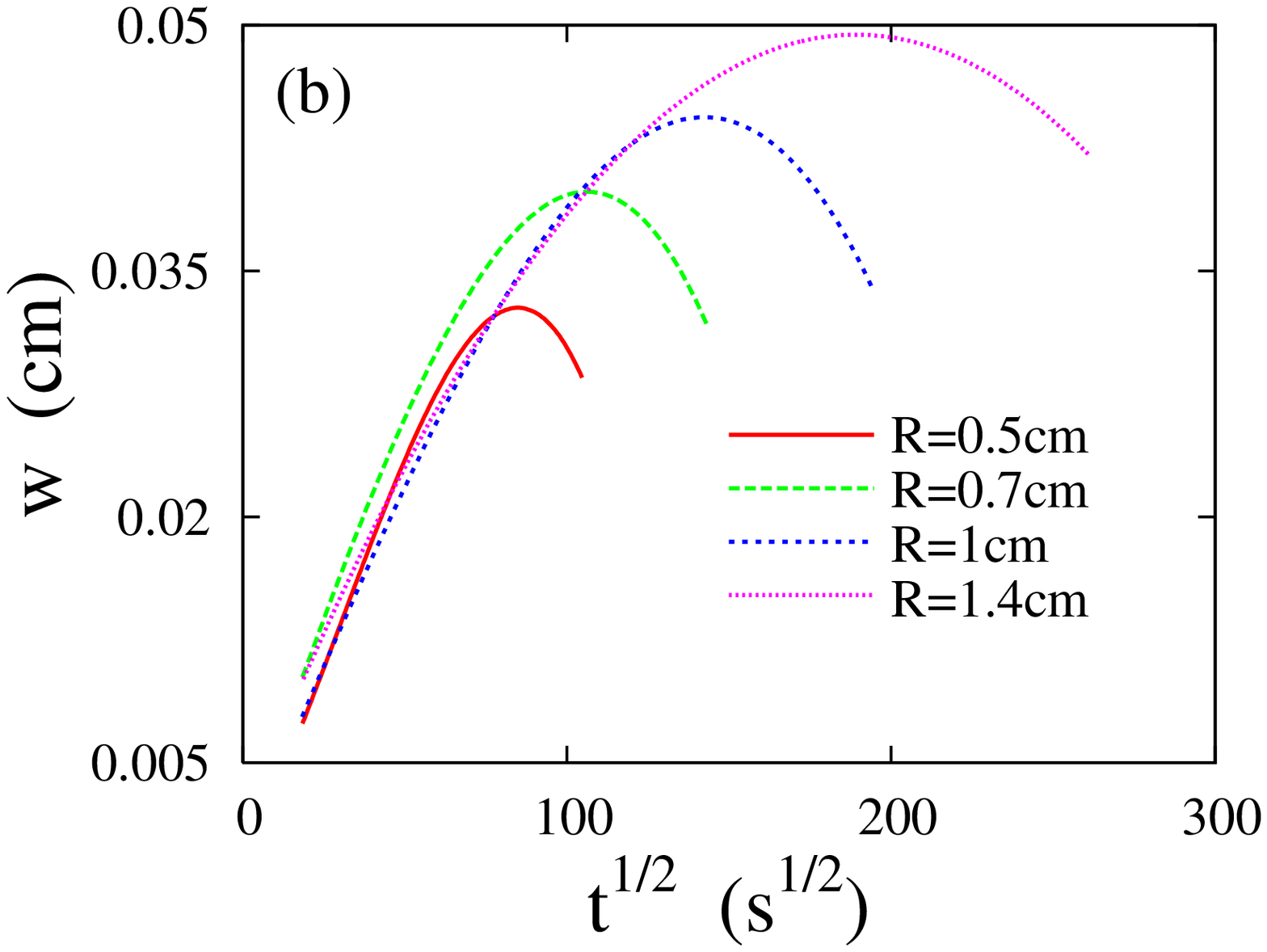}
\caption{Radius (a) and width (b) of the precipitation ring plotted
as a function of $\sqrt{t}$ with time measured in seconds. The
position of both the leading edge and the back of the ring are
plotted on (a) for various radii ($R$) of the reservoir where the
outer electrolyte is confined initially.} \label{frontelolhatul}
\end{figure}

The width $w(t)$ of the pulse shows a more complicated behavior
(Figure  1b). It increases up to a maximum value and, as the front
significantly slows, it starts to decrease.  In this  regime, the
precipitation stops at the outer edge of the pattern, the reason
being that the  limited  feeding cannot  maintain the necessary
concentration for the  formation of the precipitate.  Nevertheless,
even the limited feeding is sufficient for the complex formation in
the wake of the front, and $w$ decreases.

\begin{figure}[ht!]
\includegraphics[width=8.0cm]{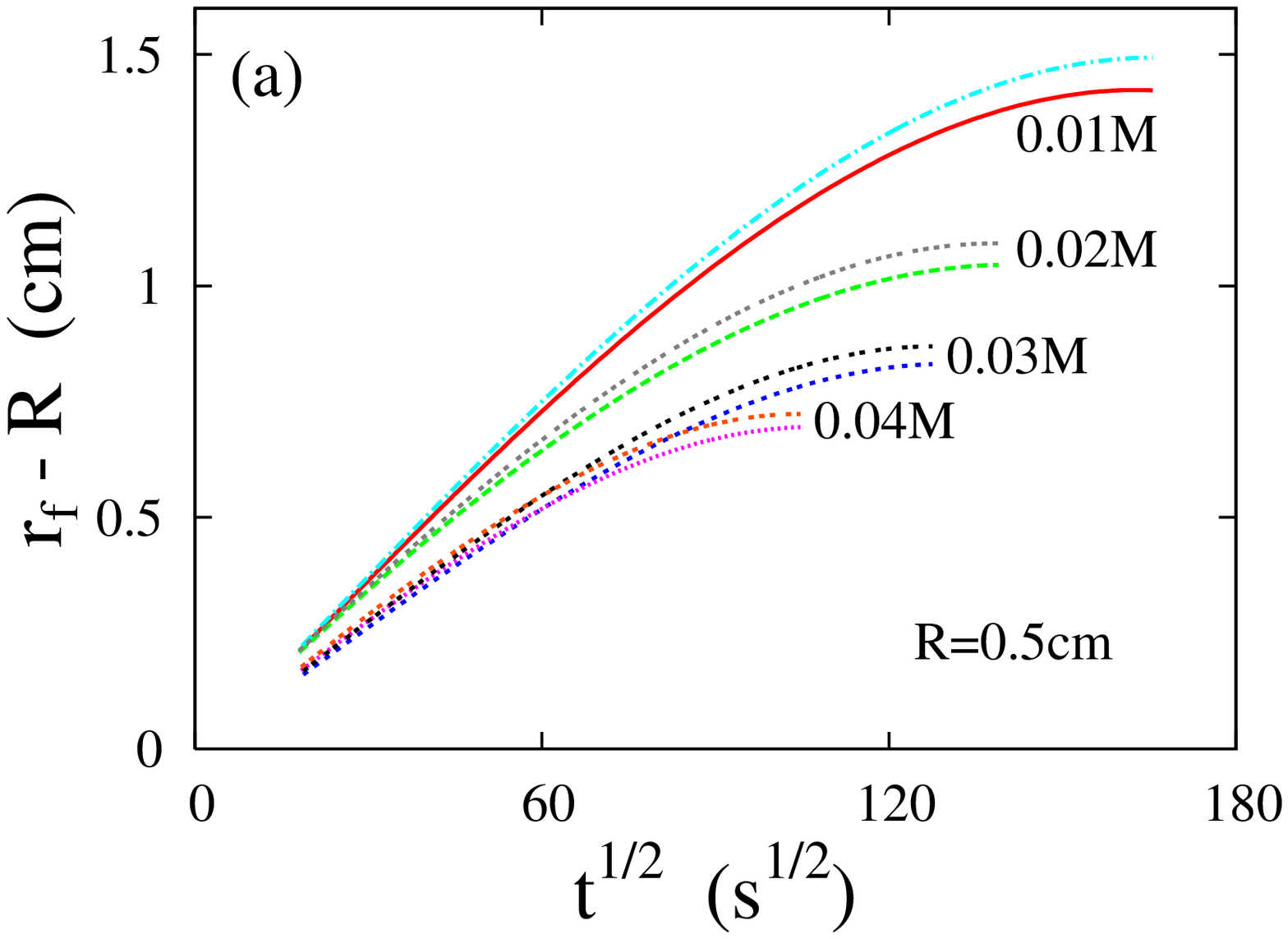}
\includegraphics[width=8.0cm]{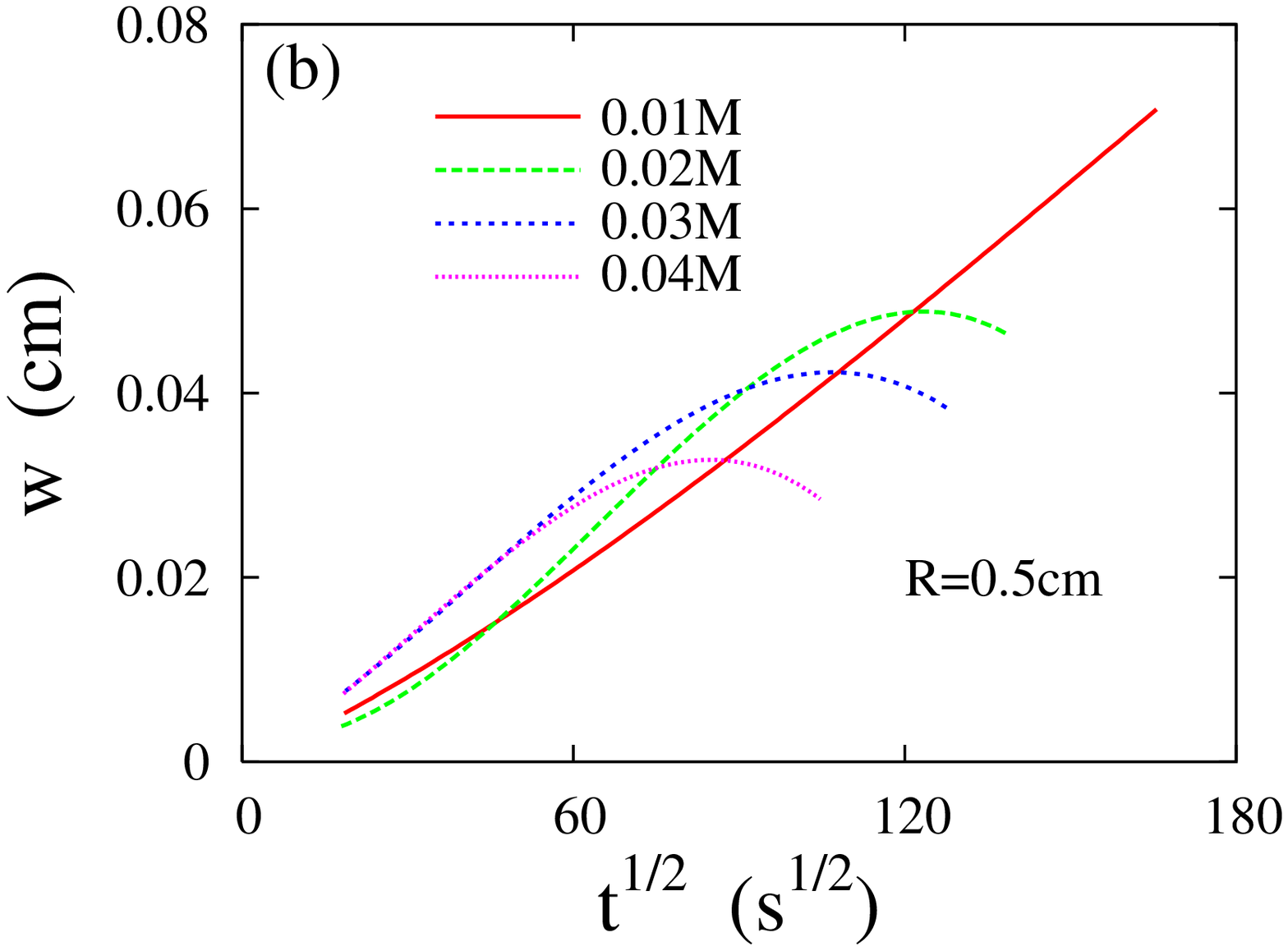}
\caption{The same as on Fig.1 a and b but for various initial
concentration of the inner electrolyte. The radius of the container
where the outer electrolyte is initially confined is also kept
constant ($R=0.5cm$).} \label{c_altalanos}
\end{figure}

Similar sequence of events
can  be observed at fixed $R$ (fixed amount of  the outer
electrolyte) when varying  the concentration of the  inner
electrolyte  (Figure  2). The effect of concentration of the
electrolytes on front velocity has been studied in previously
\cite{Zrinyi91,Sultan02}, and it was found that the velocity of
pattern formation is higher at lower concentration of the inner
electrolyte. Here we find that lowering of concentration of the
inner electrolyte mainly affects the temporal evolution of the
width. Namely, for very low concentrations, $w(t)\sim \sqrt{t}$
behavior can be seen all the way to the time when the pulse stops.
We do not have an explanation for this. The observation that the
concentration of the precipitate visibly decreases at the outer edge
of the front, however, suggests that a colloidal re-transformation
of the precipitation ring compensates the complex formation which
would otherwise decrease the width.

\begin{figure*}[ht!]
\resizebox{17cm}{!}{\includegraphics{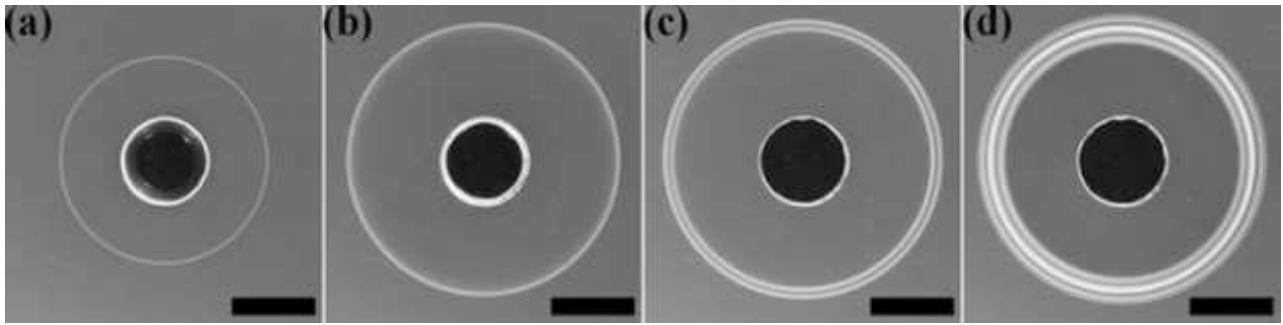}} \caption{Evolution of
the precipitation pulse ([AlCl$_3]_0$=0.02M, [NaOH]$_0$=2.50M,
$V_0=125 \mu L$, and $R=0.5$ cm): (a) single ring at $t=1h$; (b)
band stops ($t=1.5h$); (c) new ring forms inside the outer ring
($t=5h$); (d) triple ringed structure ($t=6.5h$). The scale bar is 1
cm.} \label{Exp-rev-multi}
\end{figure*}

An interesting phenomenon can be observed after the precipitation
pulse stops. As shown in Fig.3, a new well-separated ring emerges
inside the stationary pulse. It appears that the precipitation front
reverses its motion. Supporting the picture of the reversal of the
front is that a third ring with even smaller radius forms after a
while and, furthermore, the outer rings do not get dissolved. This
means that the outer electrolyte is confined to regions with radius
smaller than that of the stationary rings. It should be noted that
reversal of reaction front motion (actually several switches in the
direction of the motion) have been seen and discussed previously
\cite{Taitelbaum1,Taitelbaum2,Taitelbaum3}. The emphasis of our work
is on the motion of the precipitation pulse which is related
to the dynamics of the reaction front but it is more complex
due to the phase separation and redissolution effects.

\section{The model and the simulation results}
\label{themodel}

In order to model the phenomena described in Sec.\ref{Exp}, let us
first note that the reactions take place in a gel. Thus, no
convection is present, and the primary dynamics is a slow
reaction-diffusion process. Denoting the reagents of the outer- and
inner-electrolytes by $A\equiv$ OH$^-$ and $B\equiv $ Al$^{3+}$,
respectively, the first stage of the process, $3A+B\to C$,  yields a
reaction product $C\equiv $ Al(OH)$_3$. The next step is the
formation of precipitate which will be modeled as phase separation
of $C$ into low- and high-concentration regions of $C$ as described
by the Cahn-Hilliard equation. A similar strategy has been
successfully employed in the theory of Liesegang phenomena
\cite{ModelB,Kinis} where the above two steps constitute the whole
pattern formation process. The new element here is that the outer
electrolyte forms a complex with the precipitate ($A+C\to D\equiv$
[Al(OH)$_4]^{-}$). The complex formation is responsible for the
redissolution of $C$ and, consequently, for the moving precipitation
band. For simplicity, we shall assume that, once the complex is
formed, it ceases to play any active role in the reactions. This
assumption appears to be justified in the given experiment.

Denoting the concentrations of $A$, $B$, and $C$ by $a$, $b$, and
$c$, respectively, the above understanding (model) is described by
the following equations
\begin{eqnarray}
    \partial_t a&=&D_a\Delta a-3ka^{3}b-ga(c-c_{l})\nonumber\\
    \partial_t b&=&D_b\Delta b-ka^{3}b\label{RD-equations}\\
    \partial_t c&=&-\lambda\Delta
    \frac{\delta f}{\delta c}+ka^{3}b-ga(c-c_{l}) \, .\nonumber
\end{eqnarray}
Here $D_a$, $D_b$ are the diffusion coefficient of $A$ and $B$,
respectively, $k$ and $g$ are the reaction rate constants, $\lambda$
is the kinetic coefficient giving the timescale of the phase
separation process. The phase separation is governed by a free
energy functional $f(c)$ with two minima in homogeneous states
corresponding to the equilibrium high- ($c_{h}$) and low-
($c_l\approx 0$) concentrations of $C$ (which are assumed to be the
band- and interband-concentrations of $C$). The usual form of this
functional is the Landau-Ginzburg free energy with a functional
derivative of the following form
\begin{equation}
\frac{\delta f(c)}{\delta c}=u_0(c-\bar c)+
v_0(c-\bar c)^3+w_0\Delta c
\label{free-energy}
\end{equation}
where $u_0$, $v_0$, and $\bar c$ are phenomenological parameters
chosen so that the minima of $f(c)$ are at $c_l$ and $c_h$, while
$w_0$ governs the scale for the width of the interface between the
regions of $c_l$ and $c_h$. The values of these parameters are not known
(note that since the values of $c_l$ and $c_h$, and the width of the
interface regions are not measured, the values of $u_0$, $v_0$, and $w_0$
are not fixed in the given experiment).

According to the experiments, the outer- (inner-) electrolytes of
concentrations $a_0$ ($b_0$) are initially homogeneously distributed
inside (outside) of a circle of radius $R$.  Thus, denoting the
radial coordinate by $r$, the initial conditions to equations (1-3)
are as follows
\begin{eqnarray}
        a(r,t=0)&=&a_0\theta(R-r) \nonumber \\
        b(r,t=0)&=&b_0\theta(r-R) \label{init-cond}\\
        c(r,t=0)&=&0 \nonumber\;
\end{eqnarray}
where $\theta(x)$ is the step function. The radial symmetry is not
broken during the evolution of the precipitation pattern and,
consequently, we can limit the solution of equations (1-3) to
radially symmetric concentration fields $a(r,t)$, $b(r,t)$, and
$c(r,t)$. The reduction to a one-dimensional problem greatly
simplifies the numerical solution of the equations and even a simple
Euler scheme is sufficiently fast and stable to obtain the time
evolution of the system.

\begin{figure*}[ht!]
\includegraphics[width=4.2cm]{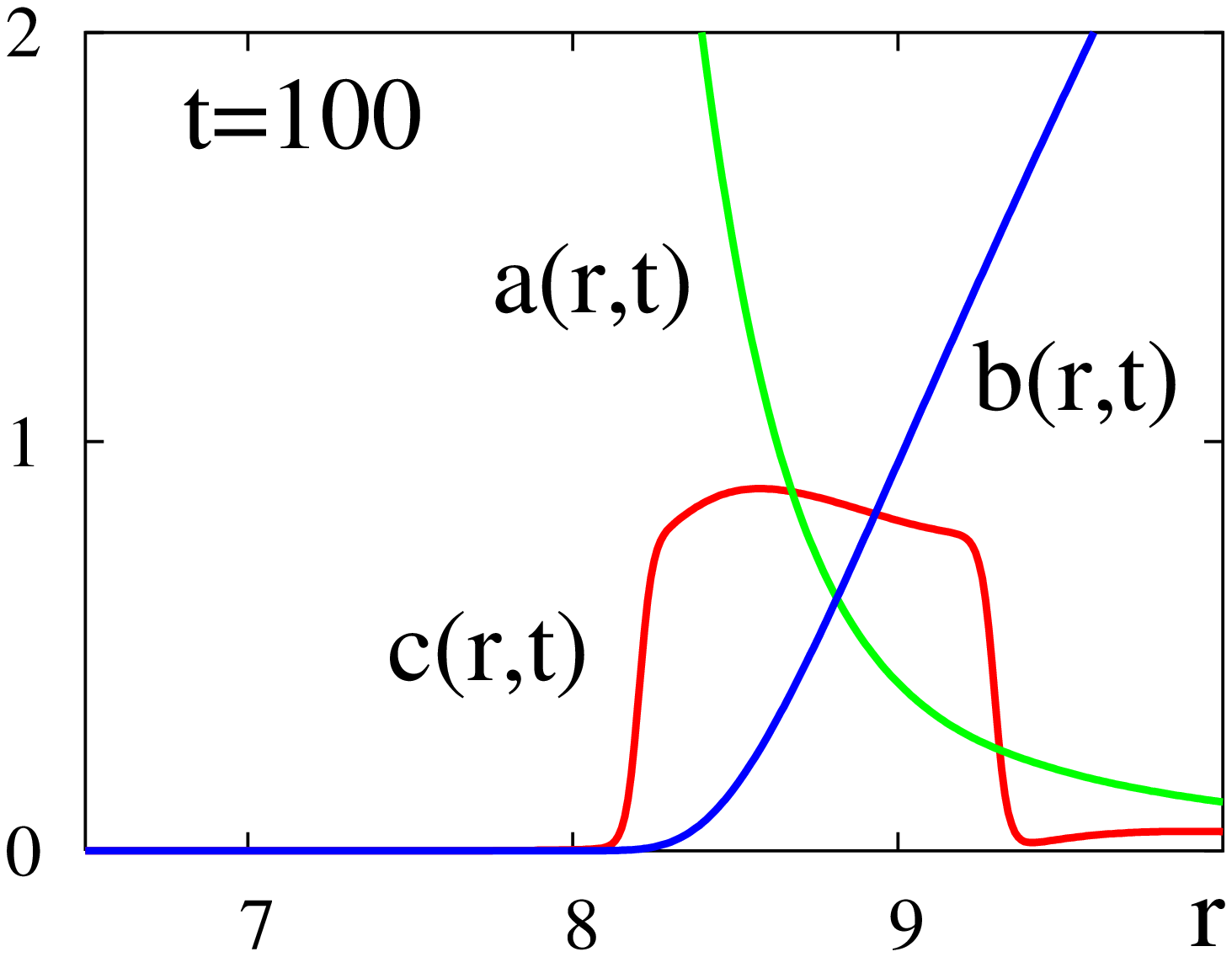}
\includegraphics[width=4.2cm]{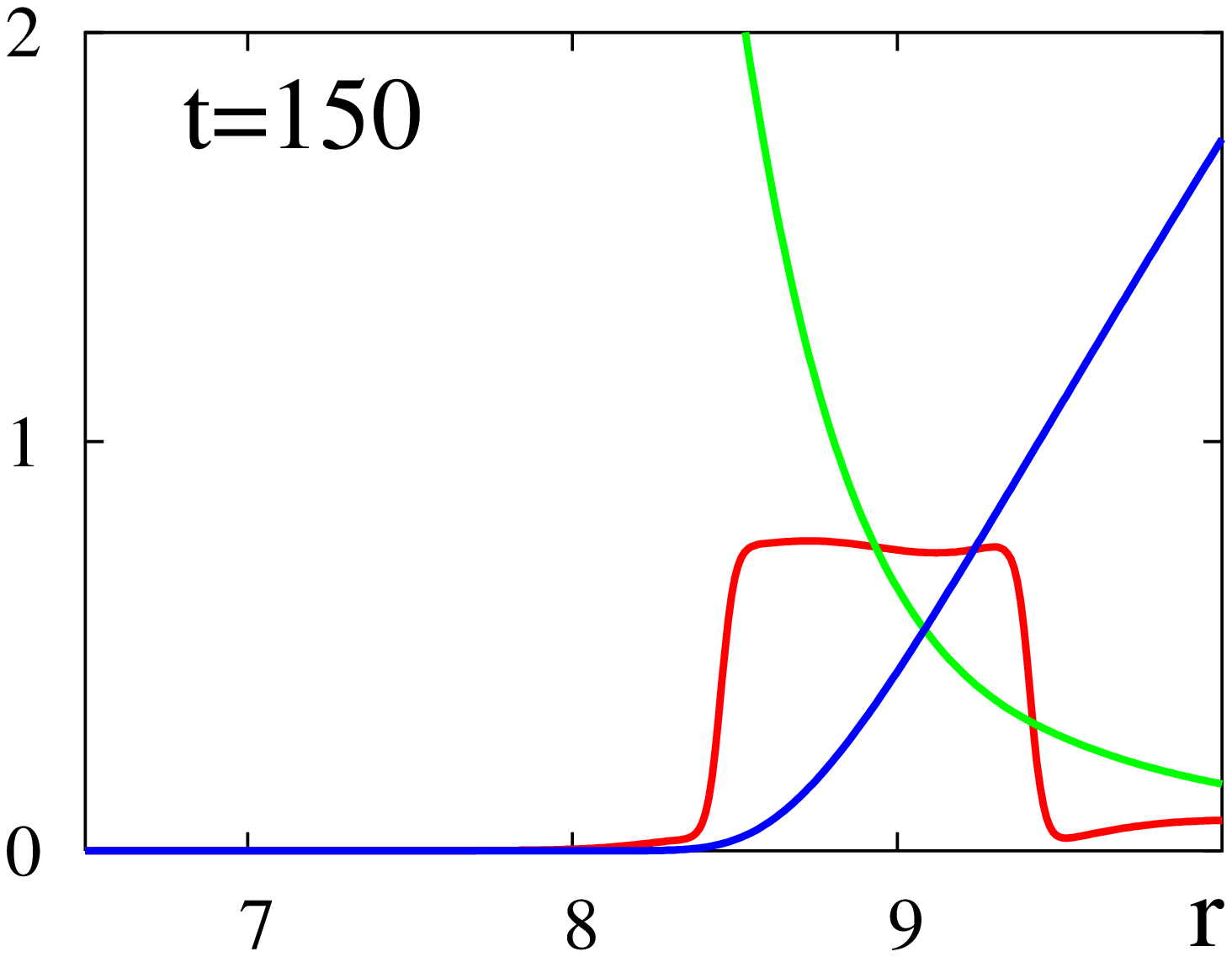}
\includegraphics[width=4.2cm]{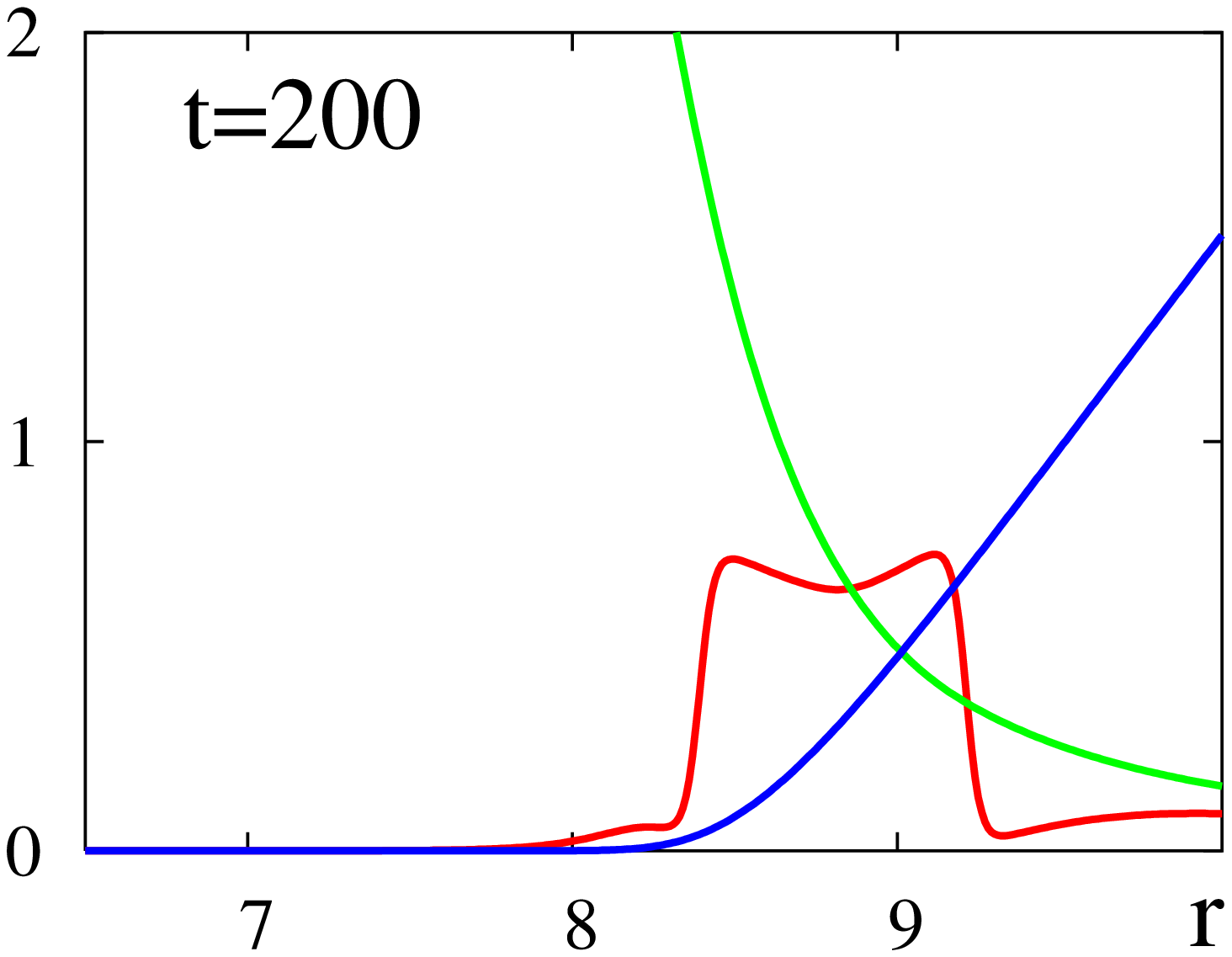}
\includegraphics[width=4.2cm]{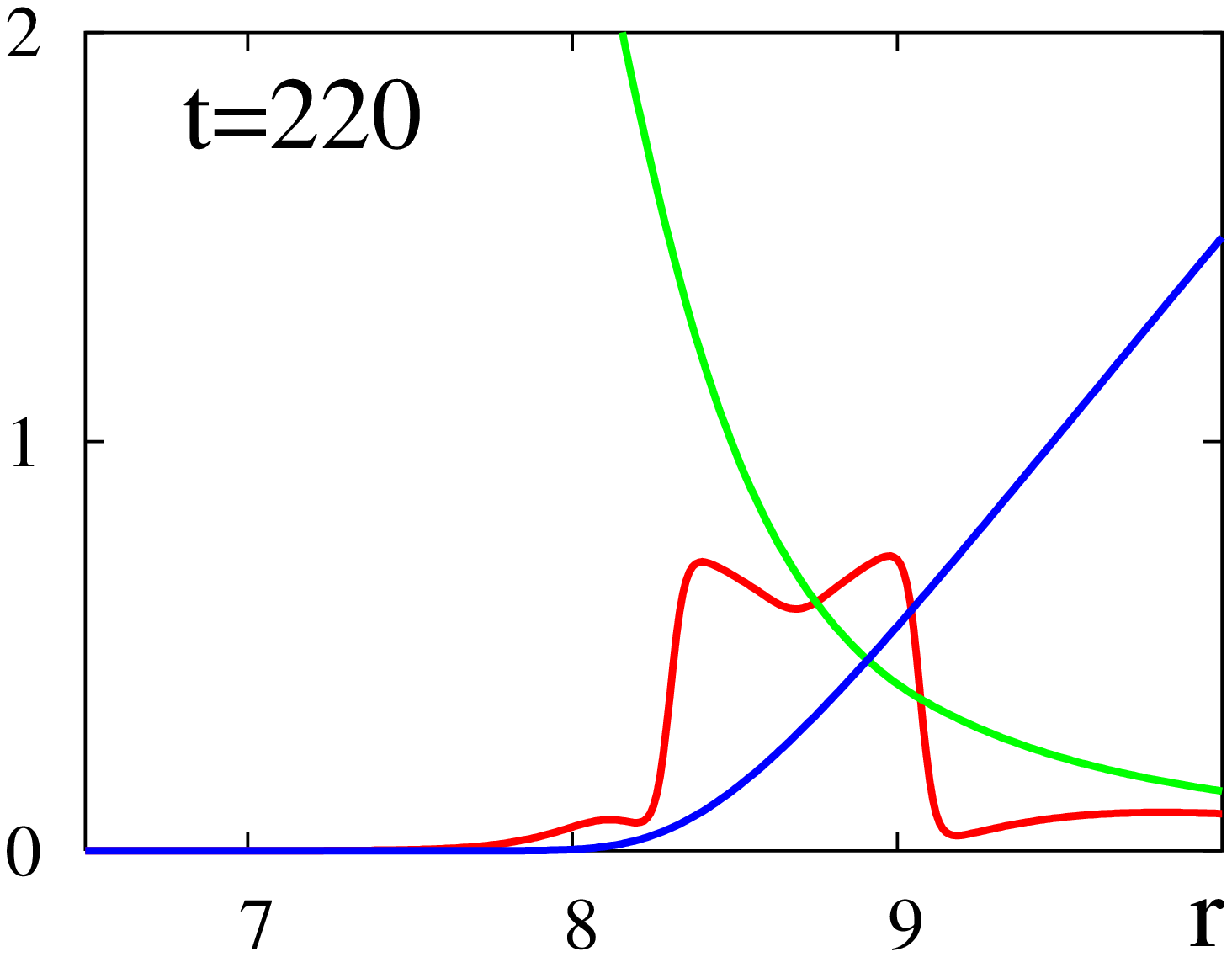}
\includegraphics[width=4.2cm]{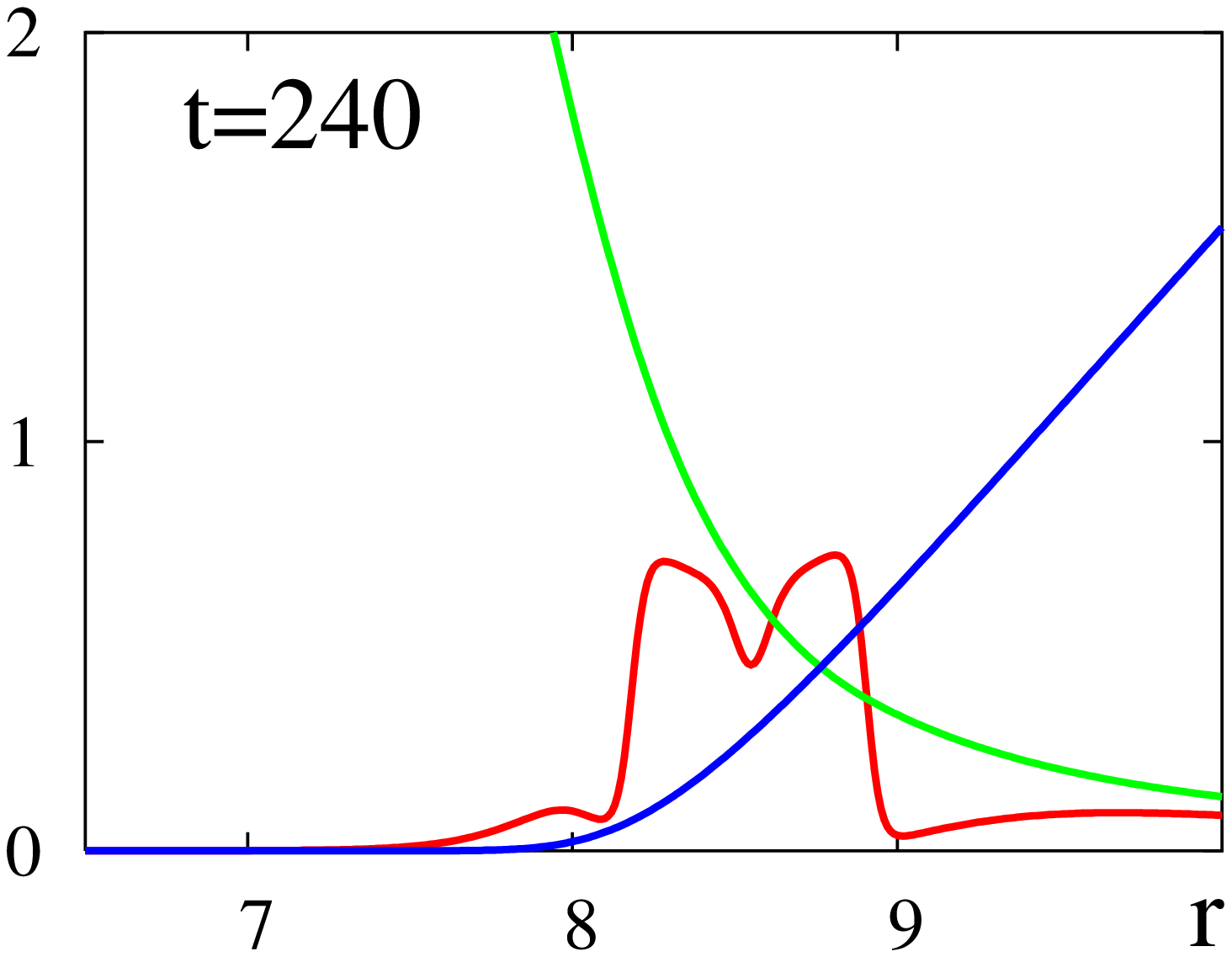}
\includegraphics[width=4.2cm]{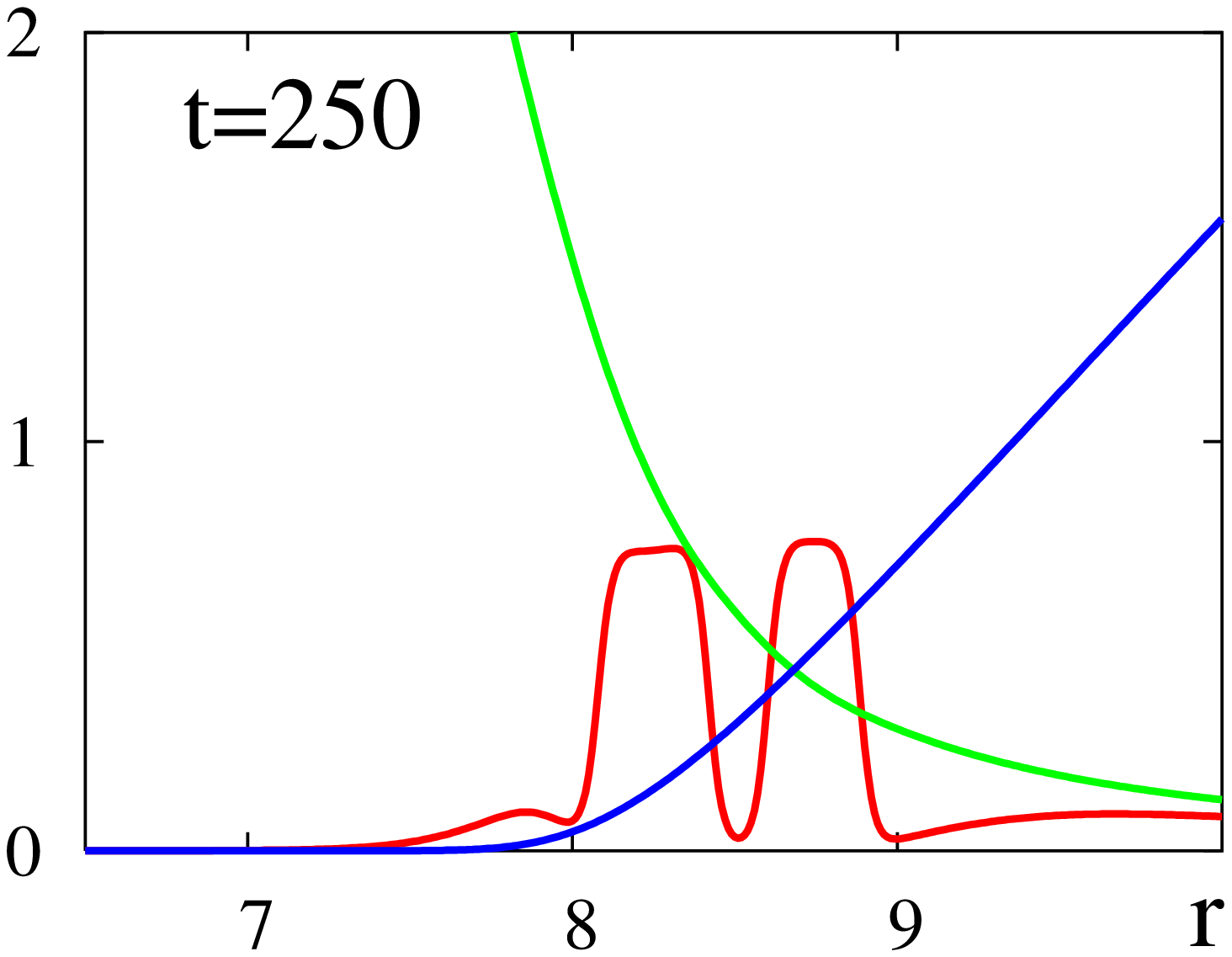}
\includegraphics[width=4.2cm]{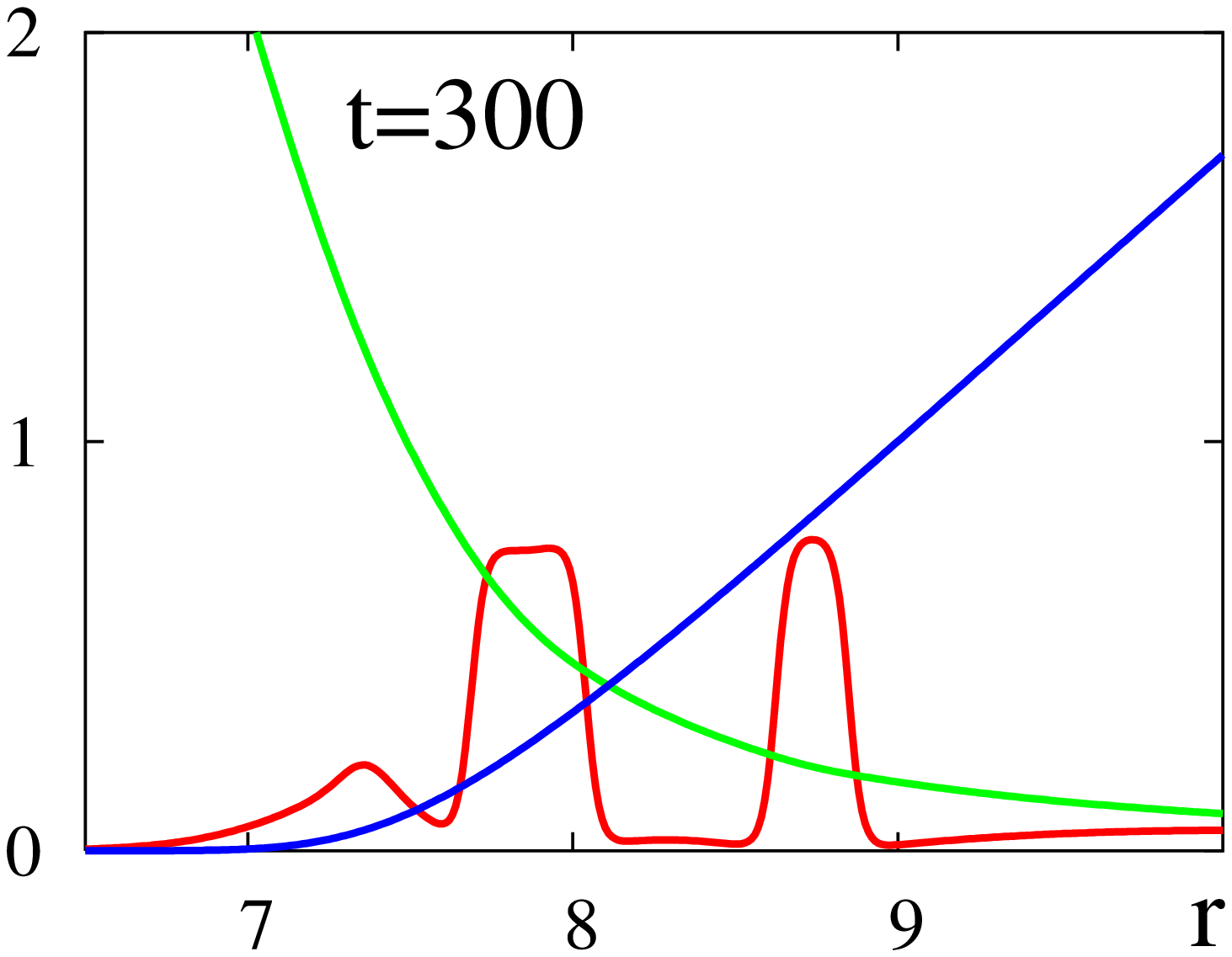}
\includegraphics[width=4.2cm]{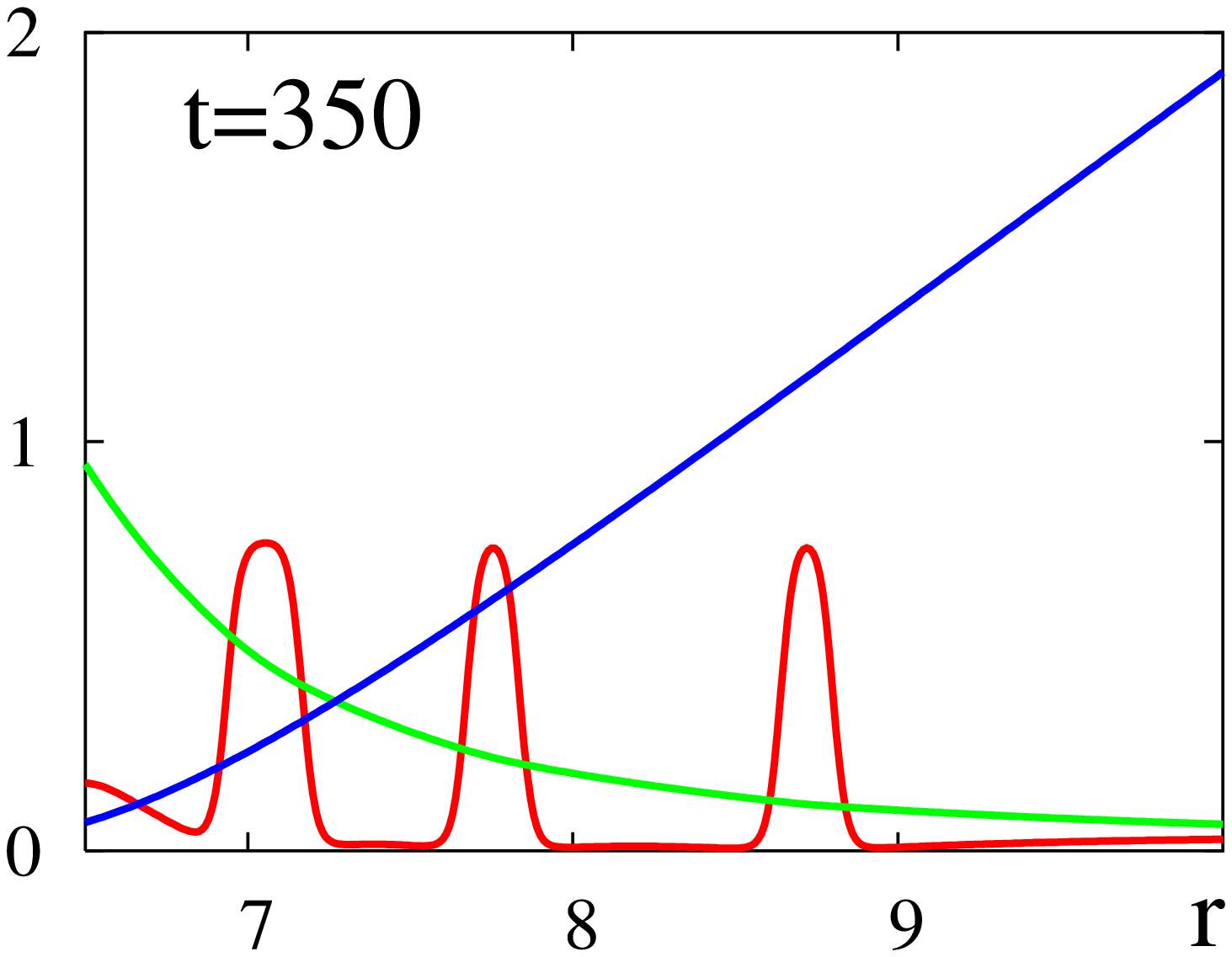}

\caption{Formation of
multiple bands after reversal of the reaction front. First panel
shows the line-assignment to the outer $(B)$ and inner $(A)$
electrolytes, and to the precipitate $(C)$. Detailed explanation of
the process can be found in the text. The above figures were
obtained by solving eqs.(\ref{RD-equations}) using the following
parameters: $a_0 = 25$, $b_0 = 0.4$, $u = 0.15$, $v = 550$, $w = 1$,
$k = 4$, $g = 0.005$, and $R=350$.} \label{Reverse-bands}
\end{figure*}

The difficulty lies in the number of phenomenological parameters.
Setting the concentration scale and the length-scale by $a_0$ and
$R$, respectively, using a timescale which yields $D_A=1$, and
making a further simplification by assuming $D_B=D_A$, one still has
five parameters ($u=\lambda u_0$, $v=\lambda v_0$, $w=\lambda w_0$,
$k$, $g$). The values of these parameters are not known and they can
change by several orders of magnitudes in various systems. Thus, one
faces the problem of searching for agreement with the experiment in
a five-dimensional parameter space. This is a highly nontrivial
problem, and we began by restricting our search to qualitative
solutions explaining the most interesting part of the experiment,
namely the reversal of the motion of front and the generation of
stationary bands in the wake of it.

Fig.\ref{Reverse-bands} shows the details of dynamics of the
reversal and the formation of multiple bands. As the outer
electrolyte $A$ is depleted, the reaction front stops and, as a
result, the outer edge of the precipitate ceases moving
($t=100-150$). Then, due to dissolution, the outer edge starts to
move backward ($t=150-200$) and the band becomes narrower. As the
reaction front reverses its motion, the precipitation takes place
behind the outer edge ($t=200-240$). During this stage, a
complicated interplay of reaction-precipitation-dissolution yields a
double peak band structure. As the reaction front moves in further,
the inner edge of the band starts to move in as well ($t=240-250$).
Now, the concentration of $A$ is still large enough  behind the
front to dissolve part of the band, and a band-splitting takes place
($t=250$). As the front moves in further, the inner band follows it
and the $A$-s are so much depleted in the wake of the front that
they are not able to dissolve the outer band ($t=250-300$). From
this point on, the process is similar to usual Liesegang band
formation \cite{ModelB}: the reaction product of the moving front
phase separates in the wake of the front, and the bands formed in
this way are stable since the dissolution is negligible due to the
depletion of the $A$-s ($t=300-350$). The band formation stops once
the concentration of the reaction product in the front decreases
below a threshold set by the Cahn-Hilliard equation.
\begin{table}[htb]
\begin{center}
\begin{tabular}{| c | c | c | c | c |}
\hline
$R$ (mm)& 5\quad & 7 \quad& 10 \quad& 14\quad \\
\hline
 $r_f(\infty)/R$& $2.44\pm 0.07$ \quad & $2.52\pm 0.05$ \quad &$2.45\pm 0.03$\quad  &$2.46\pm 0.02$ \quad \\
\hline
\end{tabular}
\end{center}
\caption{Experimentally observed relationship between the radius $R$
of the initial interface between the reagents and the radius
$r_f(\infty)$ of the final position of the precipitation ring
([AlCl$_3]_0$=0.04M).} \label{t_ter1}
\end{table}

\begin{figure}[ht!]
\includegraphics[width=8.0cm]{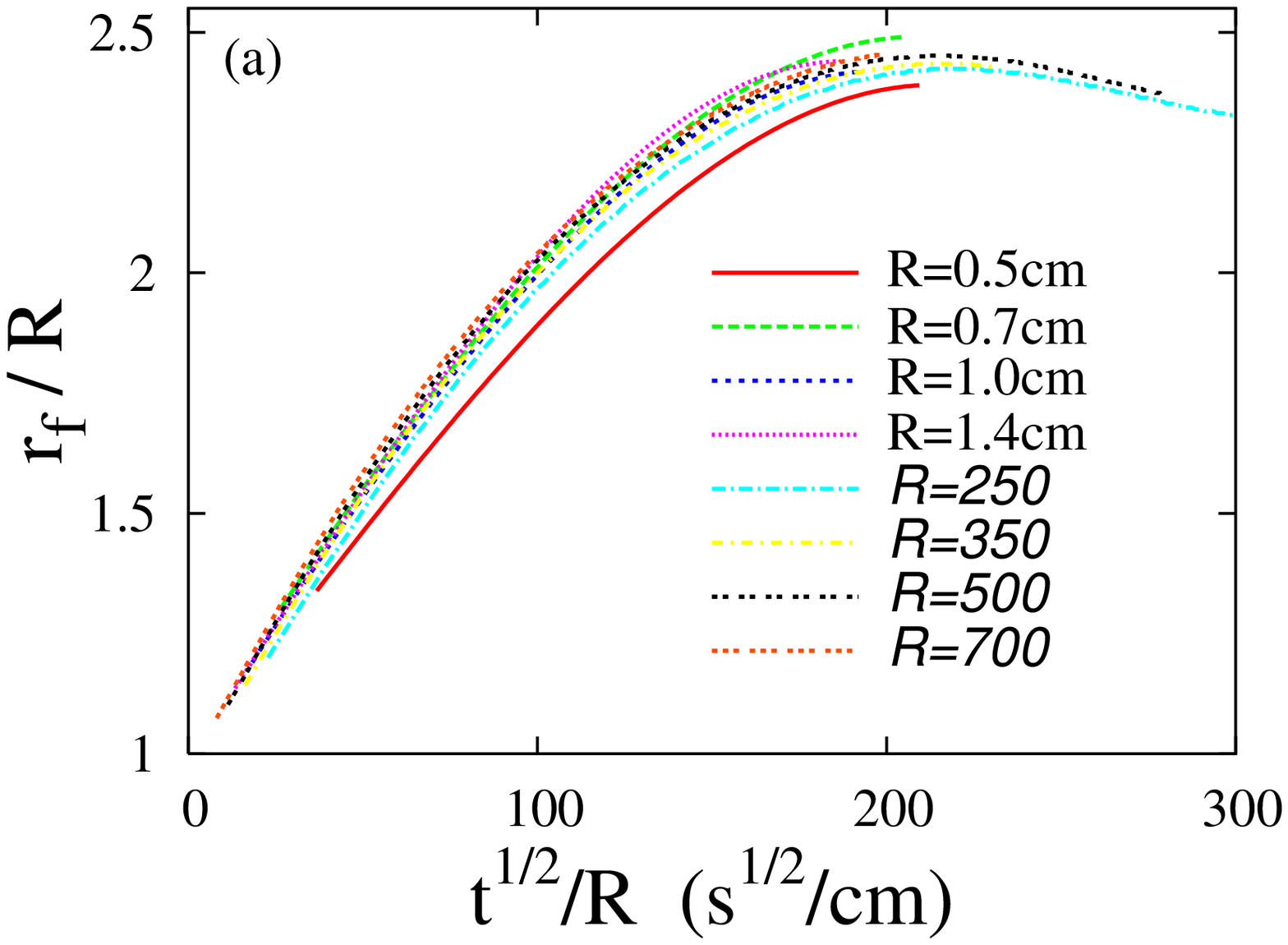}
\includegraphics[width=8.0cm]{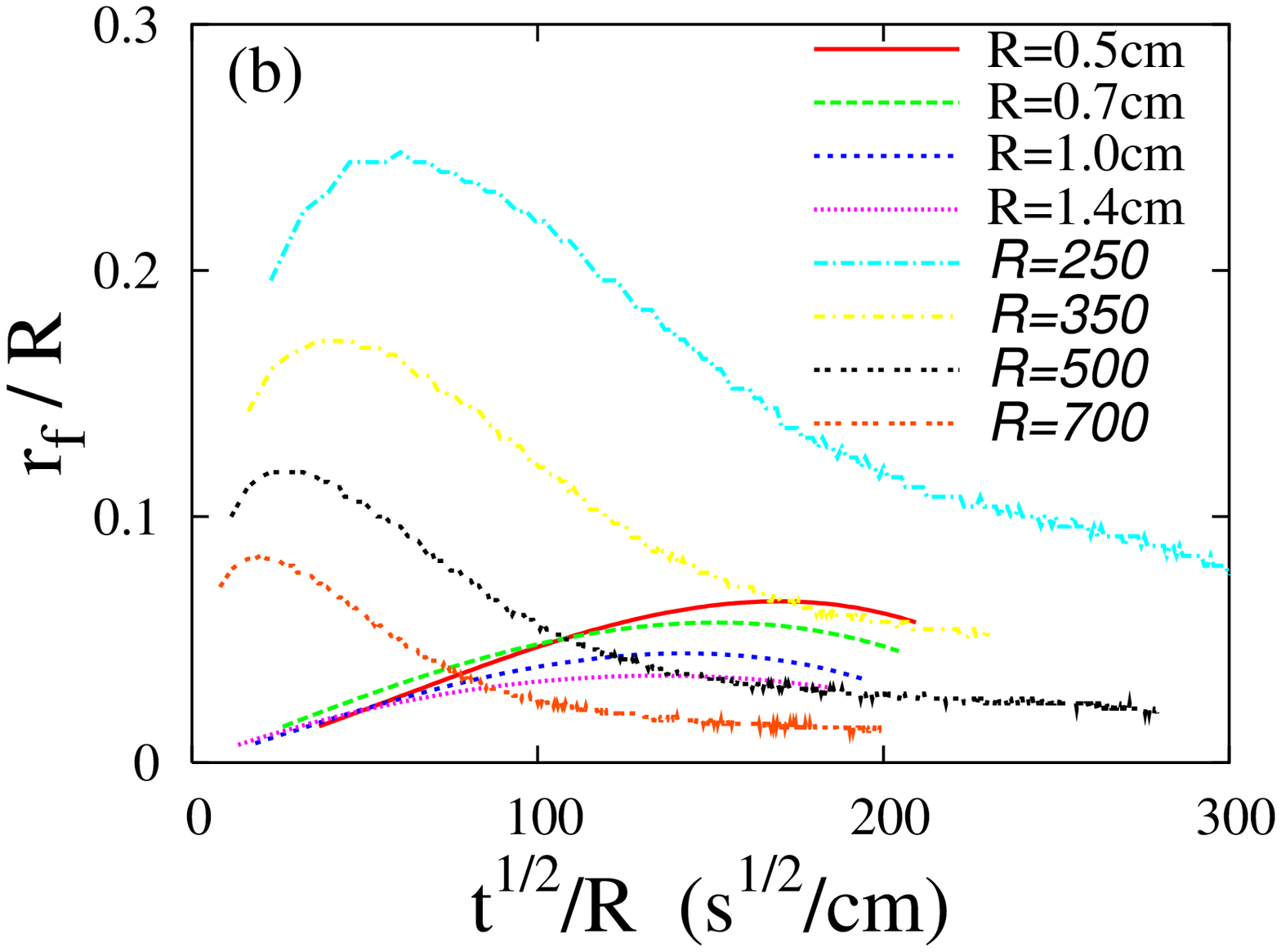}
\caption{Time evolution of the radius (a) and the width (b) of the pulse
scaled by the radius of the initial interface ($R$). The time is
scaled by $R^2$. Both experimental (R) and theoretical ($R$) results
are shown with the scales on the horizontal axis related through
$(t^{1/2}/R)_{exp}=5.7\cdot 10^3(t^{1/2}/R)_{th}$. The parameters
used in the numerical solution of
Eqs.(\ref{RD-equations}-\ref{init-cond}) were
$a_0=25$, $b_0=0.4$, $u = 0.4$, $v = 100$, $w = 1$, $k = 0.3$ and $g
= 0.002$.} \label{Front-pos-exp-num}
\end{figure}

Once qualitative agreement is found, a quantitative comparison with
the experiments may also be attempted. E.g. the time-evolution of
the radius $r_f(t)$ and width $w(t)$ of the pulse can be calculated
and measured in detail. We did this in a regime where the
concentrations of $a_0$ and $b_0$ are chosen such that the front
just driven out, then stops, and no reverse motion takes place.
Fig.\ref{Front-pos-exp-num}a shows a set of experimental and
theoretical results for time evolution of the front position,
$r_f(t)$ scaled by the radius of the initial interface ($R$). The
scaling by $R$ is suggested by calculating the final position of the
front [$r_f(\infty)$] by finding the radius of area where the $B$-s
are in sufficient number to consume all the $A$-s. This calculation
suggests that $r_f(\infty)/R$ is constant, independent of R. As can
be seen from Table 1, the relationship $r_f(\infty)/R= constant$ is
indeed satisfied.

On Fig.\ref{Front-pos-exp-num}a, the radius of the pulse is plotted
against $t^{1/2}/R$ and good collapse is found for both experimental
and theoretical curves for various $R$-s. This shows that although
the front motion is not simply diffusive, the front dynamics has the
scale-invariance of diffusive processes, and the theoretical model
captures this aspect correctly. Note that the agreement between the
experiment and theory is also excellent. The collapse is achieved
for the experimental ratio of initial concentrations
($b_0/a_0=0.016$) by finding an appropriate set of reaction rates
and Landau coefficients, and using a scale factor to relate the
experimental and theoretical timescales (see caption to
Fig.\ref{Front-pos-exp-num}).

Problems arise in the quantitative description of the width of the
precipitation zone. As can be seen on Fig.\ref{Front-pos-exp-num}b,
the agreement between theory and experiment is at most qualitative.
While both widths grow initially and then show a decreasing trend,
there are important differences. The theoretical widths are
significantly larger and, furthermore, they reach their maxima well
before the band stops [$t^{1/2}/R\approx 200$, see
Fig.\ref{Front-pos-exp-num}a]. This is in contrast to the
experimental widths which increase almost to the end of the band
motion. The narrower experimental widths may be explained by the
uncertainties of what is measured in the experiment as compared to
the theory. The shift of the maxima, however, is harder to explain.
The underlying reason may be the simplicity of the Landau free
energy where the $u_0$ and $v_0$ set a constant concentration of $C$
in the band. Visual observations, however, indicate that the
structure of the band and thus the concentration in them is changing
during the evolution. Consequently, $u_0$ and $v_0$, and perhaps
$\lambda$ and $w_0$ as well, are functions of the local
concentration of $A$.

The $a$-dependence of the free energy is an interesting problem and
the solution may also be relevant for building a more powerful
theory of Liesegang phenomena \cite{Liese-width1,Liese-width2}.
Indeed, the explanation of the so called width-law of Liesegang
phenomena is based on phase-separation scenario with an underlying
free energy. A consequence of this theory is the constancy of the
precipitate concentration in the bands which is not always seen in
experiments \cite{Zrinyi}. A natural explanation is again in the
generalization of the free energy. At this point, however, one would
need more experimental input to build a more complicated model. This
is outside the scope of this letter.
\\
\section{Final remarks}
\label{summary}

We have investigated a precipitation process where limited feeding
of the invading electrolyte, precipitation, and a complex-formation
process combine to yield a precipitation pulse with nontrivial
motion and with the possibility of evolution into a multi-band
structure. An explanation for the various phenomena can be found
through the reaction front dynamics governed by the direction of the
higher transport flux of electrolytes. In contrast to classical
setups where the high initial concentration of the outer electrolyte
ensures its dominant flux into inner electrolyte, the limited
feeding of the external electrolyte produces a reaction-front
reversal. As a consequence, a Liesegang type band-formation is
developed in direction of the source of the external electrolyte.
Presumably, this type of reaction-front reversals underlie the
explanation of several seemingly revert Liesegang structures in
chemistry and geoscience. The experimental results have been
described using a phase separation scenario for the precipitation
formation and adding the process of redissolution through complex
formation. The model captures all important features and, we
believe, it may be the basis for understanding rather complex
Liesegang type precipitation patterns.

\section{Acknowledgments}

This research has been supported by the Hungarian Academy of
Sciences (OTKA T043734 and D048673) and {\"O}veges Research
Fellowship of the National Office for Research and Technology.

\end{document}